\documentclass[12pt]{article}

\usepackage{setspace}
\usepackage{bigints}
\usepackage{amsfonts}
\usepackage{amsthm}
\usepackage{amsmath}
\usepackage{graphicx}
\usepackage{psfrag}
\usepackage{amssymb}
\usepackage{tikz}
\usepackage{authblk}
\usetikzlibrary{arrows.meta}
\usepackage[margin=2.5cm]{geometry}
\usepackage{etoolbox}
\makeatletter
\pretocmd\@maketitle
  {\def\@makefnmark{\hbox{\@textsuperscript{\normalfont\@thefnmark}}}}
\makeatother

\swapnumbers

\newtheorem*{Thm2.1.3}{{Theorem 2.1.3 of \cite{CoddingtonTheory}}}
\newtheorem*{ThmI.5.3}{{Theorem I.5.3 of \cite{Hale2009Ordinary}}}

\theoremstyle{definition}

\theoremstyle{remark}

\newcommand{\man}[1]{\ensuremath{\mathcal{#1}}} 

\newcommand{\R}{\ensuremath{\mathbb{R}}}
\newcommand{\N}{\ensuremath{\mathbb{N}}}

\begin{document}

  \title{\bf What actually happens when you approach a gravitational singularity?\thanks{Essay written for the Gravity Research Foundation 2021 Awards for Essays on Gravitation.}}
  \author{Susan M Scott\thanks{Corresponding author}\thanks{Email address: susan.scott@anu.edu.au} and Ben E Whale\thanks{Email address: ben@benwhale.com}\\
  {\small\it Centre for Gravitational Astrophysics\\
  \it Research School of Physics\\
  \it College of Science\\
  \it The Australian National University\\
  \it Building 38A, Science Road\\
  \it Acton  ACT  2601\\
  \it Australia}
  }



  \maketitle

  \begin{abstract}
    Roger Penrose's 2020 Nobel Prize in Physics recognises that his
    identification of the concepts of ``gravitational singularity'' and an
    ``incomplete, inextendible, null geodesic'' is physically very important. The
    existence of an incomplete, inextendible, null geodesic doesn't say much,
    however, if anything, about curvature divergence, nor is it a helpful
    definition for performing actual calculations. Physicists have long sought
    for a coordinate independent method of defining where a singularity is
    located, given an incomplete, inextendible, null geodesic, that also allows
    for standard analytic techniques to be implemented. In this essay we
    present a solution to this issue. It is now possible to give a concrete
    relationship between an incomplete, inextendible, null geodesic and a
    gravitational singularity, and to study any possible curvature divergence
    using standard techniques.
  \end{abstract}

  \newpage
  \setstretch{1.5}

  Roger Penrose was awarded a $50\%$ share of the 2020 Nobel Prize in Physics.
  The short form of Penrose's contribution is cited as ``for the discovery that
  black hole formation is a robust prediction of the general theory of
  relativity.'' The longer form explicitly mentions Penrose's 1965 paper that
  proved the first ``modern'' singularity theorem
  \cite{penrose1965gravitational}.

  Readers of Penrose's paper might be surprised to discover that he doesn't
  actually prove the existence of a black hole. Instead Penrose presents a very
  general result that he uses as evidence for the stability of black hole
  solutions under perturbations, specifically the Schwarzschild and Kerr black
  holes. It was the very general nature of Penrose's paper that changed the
  astronomical community's thinking about the physical reality of black holes
  and, subsequently, justified his Nobel Prize. His paper doesn't prove the
  existence of black holes, however.

  A black hole is popularly imagined as a ``place'' where gravity becomes
  infinitely strong. There are a handful of different ways to mathematically
  express this. They all boil down to the idea that the curvature of
  spacetime diverges to infinity along some curve that is assumed to approach
  the singularity. Penrose's paper says nothing about these ideas.

  Penrose proved that, under very general and physically reasonable hypotheses,
  there exists an incomplete, inextendible, null geodesic. More recent
  singularity theorems prove the existence of incomplete, inextendible,
  timelike geodesics under similarly general hypotheses. The existence of an
  incomplete, inextendible, causal geodesic challenges our physical intuition
  of time. If I have experienced a finite amount of time, then I should always
  be able to experience a bit more time. Yet the singularity theorems say that
  this is not true.

  Penrose implicitly assumes that the existence of a gravitational singularity
  is equivalent to the existence of an incomplete, inextendible, null curve. We
  agree with Penrose that a gravitational singularity is only a singularity if
  it can be approached in finite proper time. There are examples of spacetimes
  with incomplete, inextendible, timelike curves yet all causal geodesics are
  complete. Therefore, following Penrose (and ignoring issues associated with
  trapped causal curves), we take the existence of an incomplete, inextendible,
  causal curve as equivalent to the existence of a gravitational singularity.

  This is an extremely general definition of gravitational singularity, but it
  covers all possible unphysical behaviour. Now that we have an, admittedly
  difficult to handle, definition of gravitational singularity, the obvious
  question is ``Does our physical intuition fit the mathematics?''. Does curvature
  necessarily diverge along the incomplete, inextendible curves predicted by
  Penrose's singularity theorem? This question is surprisingly deep, 
  and has generated numerous branching research programs
  interested in various special cases. 

  We know a lot about what happens as an observer approaches singularities in
  certain parametrised families of spacetimes, but almost nothing about what
  happens on approach to a general singularity. This boils down to the
  difficulty of using our definition of a singularity in computations.
  This essay presents the evidence that something called, The Abstract Boundary,
  provides the tools necessary to perform curvature
  computations in general spacetimes
  starting only with the assumptions of Penrose's singularity theorem.

\begin{figure}[!hp]
    {
    \hfill
    \includegraphics[scale=0.9]{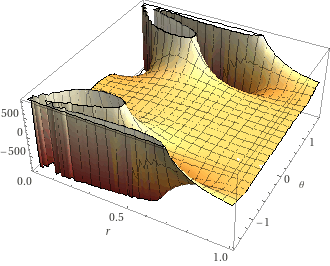}
    \hfill
    \includegraphics[scale=0.9]{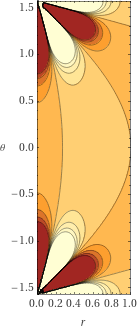}
    \hfill
    }
    \caption{\footnotesize The curvature scalar $R_{abcd}R^{abcd}$ for the Kerr black hole
      in the $(r, \theta)$ plane of Boyer-Lindquist coordinates
      \cite{HawkingEllis1973}. The left hand plot gives a 3D image, while the
      right hand plot presents the contours of the 3D image (darker colours
      represent larger values). We have truncated the value of $R_{abcd}R^{abcd}$
      at $\pm1000$. Note the dependence on $\theta$ for the value of the limit
      to $r=0$.}
    \label{fig_scalar_curvature}
  \end{figure}

  The Kerr black hole solution ($m>a$) is a clear example of how the
  analysis of a
  singularity usually proceeds. Presented in Boyer-Lindquist
  coordinates the spacetime is mapped to points in Euclidean space $\R^4$ minus
  the origin. The origin is a point on the closure of the image of the
  coordinates, however. We will call such a point a {\em boundary point}. With
  respect to this chart any incomplete, inextendible curve ends at the origin.
  It therefore appears that all incomplete, inextendible curves have the same
  endpoint at the same gravitational singularity; at the ``centre'' of the
  black hole. The curvature scalar $R_{abcd}R^{abcd}$ evaluated along
  incomplete, inextendible curves in the limit to the origin can be either
  finite or infinite. 
  See Figure (\ref{fig_scalar_curvature}).
  We call such boundary points {\em mixed}.
  
\begin{figure}[!hp]
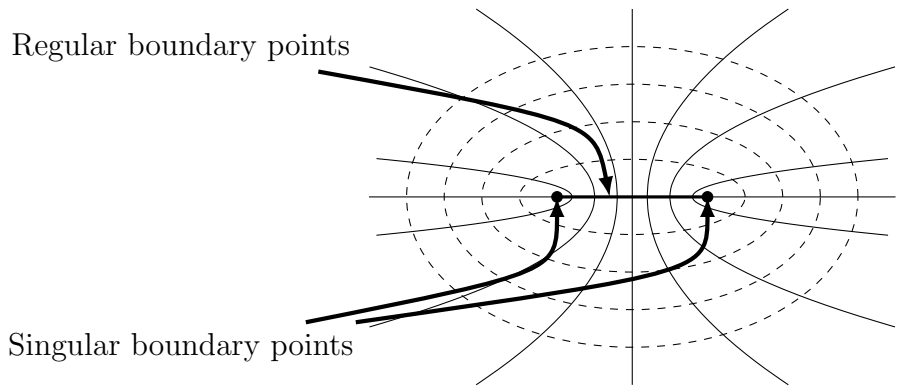

    \[
      \tikz[>=latex]{
        \foreach \o in {0} {
          \draw (-3.5+\o, 0) -- (3.5+\o, 0); 
          \draw (\o, -2.5) -- (\o, 2.5); 
          \draw[very thick] (-1+\o,0) -- (1+\o,0); 
          \foreach \c in {-1,1} { 
            \draw[fill] (\c+\o, 0) circle (2pt);
          }
          \foreach \x/\y in {1.5cm/0.5cm, 2cm/1cm, 2.5cm/1.5cm, 3cm/2cm} {
            \draw[dashed] (\o,0) circle [x radius=\x, y radius=\y]; 
          }
          \draw[domain=-1.73:1.73, smooth, variable=\y] plot ({\y*\y + 0.5 + \o}, {\y});
          \draw[domain=-1.73:1.73, smooth, variable=\y] plot ({-\y*\y - 0.5 + \o}, {\y});
          \draw[domain=-2.5:2.5, smooth, variable=\y] plot ({0.3*\y*\y + 0.2 + \o}, {\y});
          \draw[domain=-2.5:2.5, smooth, variable=\y] plot ({-0.3*\y*\y - 0.2 + \o}, {\y});
          \draw[domain=-0.51:0.51, smooth, variable=\y] plot ({10*\y*\y + 0.8 + \o}, {\y});
          \draw[domain=-0.51:0.51, smooth, variable=\y] plot ({-10*\y*\y - 0.8 +\o}, {\y});
        };

        \node at (-6, 2) (r) {Regular boundary points};
        \draw[->, ultra thick] (r) ..controls(-0.5, 1).. (-0.3, 0);
        \node at (-6, -2) (s) {Singular boundary points};
        \draw[->, ultra thick] (s) ..controls(-1, -1).. (-1,0);
        \draw[->, ultra thick] (s) ..controls(1, -1).. (1,0);
      }
    \]
    \caption{\footnotesize A profile view of the disc of boundary points associated with
      Kerr-Schild coordinates \cite{HawkingEllis1973}.
      The thick horizontal line between the
      two black dots is the set of regular boundary points 
      through which the metric can be
      extended. 
      The two black dots are the singular boundary points at which 
      $R_{abcd}R^{abcd}$ diverges.
      The dashed and solid contours are lines of constant radial and
      one of the angular Boyer-Lindquist coordinates.
      }
    \label{fig_ks}
  \end{figure}

  Switching to Kerr-Schild coordinates presents a different picture. In these
  coordinates there is a $2$-dimensional disc of boundary points. The endpoints
  of images of incomplete, inextendible, causal curves in Kerr-Schild
  coordinates are points on this disc. These curves,
  which in Boyer-Lindquist coordinates all appeared to have the same
  endpoint, could now have different endpoints. 
  See Figure (\ref{fig_ks}).
  It turns out that it is possible
  to analytically extend Kerr-Schild coordinates through the interior of the
  disc. Such boundary points are called {\em regular} as we can extend both the
  manifold and metric through them. 

  The boundary points, in Kerr-Schild coordinates, on the boundary of the disc
  are also endpoints of incomplete, inextendible curves. The curvature scalar
  $R_{abcd}R^{abcd}$ diverges to infinity, however, along all such curves, at
  these boundary points. Thus these curves can be taken to correspond to
  ``real'' gravitational singularities. We call such singularities {\em
  essential}.

  Note that in Boyer-Lindquist coordinates the internal structure of the
  singularity is muddled up with regular behaviour of the metric. Kerr-Schild
  coordinates present the structure of the singularity more cleanly. The
  physical behaviour is clearer. There are no mixed boundary points any more.
  This property of Kerr-Schild coordinates is used to justify the ring
  representation of the gravitational singularity as ``more'' physically
  reasonable than the point representation in Boyer-Lindquist
  coordinates.

  The Kerr-Schild coordinates are therefore used to give a definition of a
  gravitational singularity that is useful for computations. The union of the
  manifold with the boundary points is a closed topological space. It is
  possible to perform computations ``on the boundary'' by taking limits on
  approach to boundary points. There is a classification of boundary points
  into singularities and regular points, and we know that singular boundary
  points and incomplete, inextendible, causal curves are related.

  This definition is, however, dependent on the Kerr-Schild coordinates to
  define the boundary points. The identification of boundary points with the
  concept of gravitational singularity is chart dependent. This is a problem as
  any concept defined with respect to a manifold should be independent of
  coordinates. Moreover, it is not clear how to extend the example above to any
  Lorentzian manifold to which Penrose's singularity theorem applies. 

  Our very recent paper \cite{Scott_2021} completes a body of work
  \cite{ScottSzekeres1994,   barry2011attached, barry2014strongly,
  whale2015generalizations, Whale2014Chart,
  Ashley2002b, FamaScott1994, FamaClarke1998} that presents
  a solution to all these issues---the Abstract Boundary. The Abstract Boundary
  is an algorithm that takes a spacetime and produces a compact, non-Hausdorff,
  locally Euclidean topological space into which the original spacetime is
  homeomorphically embedded. The boundary of this embedding, also called the
  Abstract Boundary, is denoted by $\mathcal{B}(\man{M})$.

  Let $(\man{M},g)$ be a spacetime. Let $\Phi=\{\phi:\man{M} \to \man{N}\}$ be
  the set of all open embeddings of $\man{M}$ into a manifold $\man{N}$ of the
  same dimension. An element of $\mathcal{B}(\man{M})$ is an equivalence class
  of subsets of $\partial\phi(\man{M})$ for all $\phi\in\Phi$. The equivalence
  relation is defined so that for each subset,
  $A\subset\partial\phi_1(\man{M})$ and $B\subset\partial\phi_2(\man{M})$,
  $[A]=[B]\in\mathcal{B}(\man{M})$ if the set of sequences
  $(a_i)_{i\in\N}\subset \man{M}$ so that $\phi_1(a_i)\to a\in A$ and the set
  of sequences $(b_i)_{i\in\N}\subset \man{M}$ so that $\phi_2(b_i)\to b\in B$
  are the same. Full details of the construction can be found in
  \cite{ScottSzekeres1994}.
  
  In the Kerr solution, the origin in Boyer-Lindquist coordinates is equivalent
  to the disc in Kerr-Schild coordinates. See \cite{Whale2014Chart} for the
  creation of the necessary embedding from a chart. Note that the construction
  of $\mathcal{B}(\man{M})$ requires nothing more than $\man{M}$ itself. Thus
  the Abstract Boundary can be applied to any pseudo-Riemannian manifold.

\begin{figure}[!h]
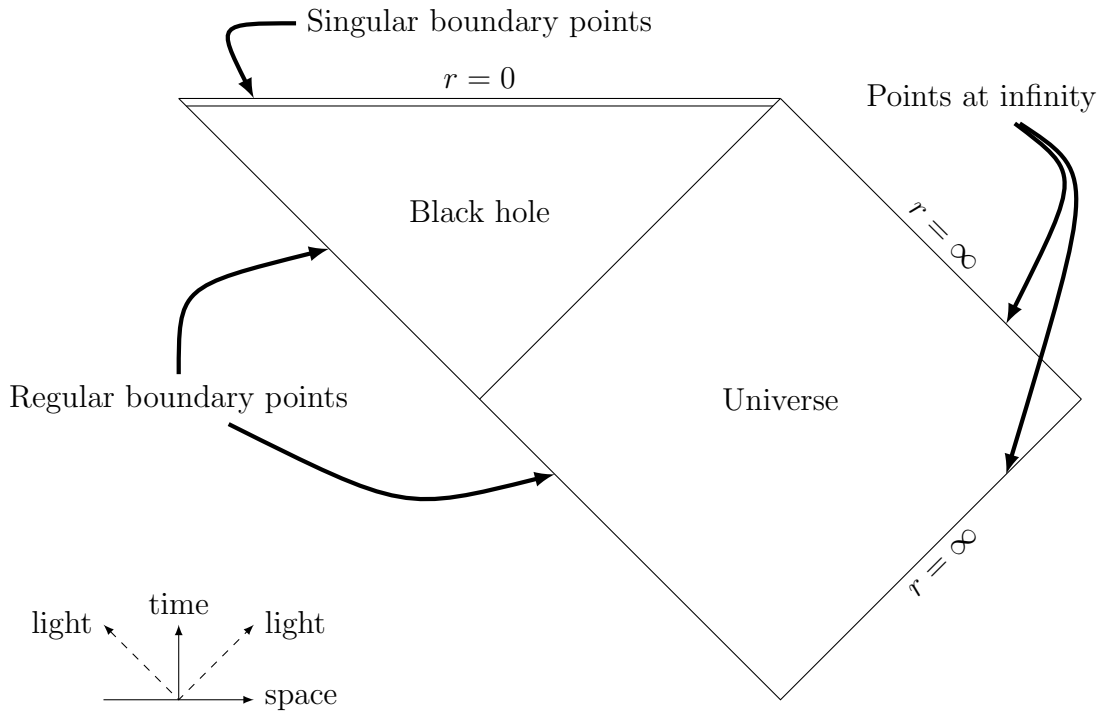

    \[
      \tikz[>=latex]{
        \draw (-4, 4) -- (4, 4) node[midway, above] {$r=0$};
        \draw (4, 4) -- (8, 0) node[midway, sloped, above] {$r=\infty$};
        \draw (8, 0) -- (4, -4) node[midway, sloped, below] {$r=\infty$} -- (-4,4);
        \draw (-3.9, 3.9) -- (3.9, 3.9);
        \draw (0,0) -- (4,4);

        \node at (-4, 0) (r) {Regular boundary points};
        \draw[->, ultra thick] (r) ..controls(-4,1.5).. (-2,2);
        \draw[->, ultra thick] (r) ..controls(-1,-1.5).. (1,-1);
points
        \node[right] at (5, 4) (i) {Points at infinity};
        \draw[->, ultra thick] (i) ..controls(8, 3).. (7, 1);
        \draw[->, ultra thick] (i) ..controls(8.2, 3).. (7, -1);

        \node at (0, 5) (s) {Singular boundary points};
        \draw[->, ultra thick] (s) ..controls(-3.5, 5).. (-3, 4);

        \node at (0, 2.5) {Black hole};
        \node at (4, 0) {Universe};

        \draw[->] (-5, -4) -- (-3, -4) node[right] {space};
        \draw[->] (-4, -4) -- (-4, -3) node[above] {time};
        \draw[<->, dashed] (-5, -3) node[above, left] {light} -- (-4, -4) -- (-3, -3) node[above, right] {light};
        
      }
    \]
        




    \caption{\footnotesize An illustration of Penrose's conformal boundary for the
      Schwarzschild black hole \cite{HawkingEllis1973}. 
      The standard classification of boundary points, in terms of the conformal
      factor, agrees with the Abstract Boundary classification 
      \cite{Whale2014Chart}.}
    \label{fig_sw}
  \end{figure}

  In this way, any method of studying a spacetime that involves attaching ideal
  points via a chart or embedding is contained in the Abstract Boundary. In
  particular, it contains, and thus generalises, both the classical coordinate
  dependent methods of studying singularities and Penrose's conformal boundary
  \cite{HawkingEllis1973} (see Figure (\ref{fig_sw})).
  Concrete examples of this can be found in \cite{Whale2014Chart}.

  Given a point $[A]$ in $\mathcal{B}(\man{M})$ we can take any representative
  $B\in[A]$. This representative is a subset of the topological boundary of
  some embedding, $B\subset\partial\phi(\man{M})$, for $\phi:\man{M}\to
  \man{N}$. One can therefore perform computations in the charts of $\man{N}$
  using the pushforward of $g$ by $\phi$. If these computations are invariant
  under the equivalence relation that defines $[A]$ then the results of the
  computation will be chart independent and well-defined. This equips the
  Abstract Boundary with a differential structure sufficient for computations
  of the sort performed with Kerr-Schild coordinates.

  One of the miracles of the Abstract Boundary is that it is possible to
  generalise the classification of boundary points, as hinted at in the Kerr
  example. Thus points in the Abstract Boundary fall into one of a handful of
  classes, including ``singularities'', ``points at infinity'' and ``mixed
  points'' \cite{ScottSzekeres1994}. The classification is constructed from the
  metric $g$, the set $\Phi$ and a set of curves (usually the set of all
  piecewise $C^1$ causal curves).

  Our recent result, the Endpoint Theorem \cite{Scott_2021}, shows that for any
  sequence $(x_i)_{i\in\N}\subset \man{M}$ with no accumulation points, there
  exists an element $[x]\in\mathcal{B}(\man{M})$ and an embedding $\phi$ so
  that $\phi(x_i)\to x$. This theorem establishes that
  $\man{M}\cup\mathcal{B}(\man{M})$ is compact in the topology given in
  \cite{barry2011attached, barry2014strongly} or \cite{Whale2014Chart} and
  thereby shows that
  $\mathcal{B}(\man{M})$ is large enough to capture all possible singular
  behaviour. Most importantly, it provides locations (the endpoints) for the
  gravitational singularities associated with the incomplete, inextendible,
  causal geodesics predicted by the singularity theorems.

  The Abstract Boundary has its own singularity theorem \cite{whale2015generalizations}, which complements Penrose's singularity theorem. The Abstract Boundary result shows that the incomplete, inextendible curves predicted by Penrose's singularity theorem have actual singularities, according to the Abstract Boundary classification, as endpoints. This renders the Abstract Boundary an excellent mathematical representation of Penrose's implicit definition of a gravitational singularity. Put another way, the Abstract Boundary is a mathematical encoding of our physical intuition.

  Lastly, the papers \cite{FamaScott1994, FamaClarke1998} show that elements of $\mathcal{B}(\man{M})$ have good topological properties under continuous maps, and the paper \cite{Ashley2002b} shows that the classification of singularities of $\mathcal{B}(\man{M})$ is stable and therefore a physically reasonable definition.
  
  The Abstract Boundary generalises the Kerr-Schild example to all Lorentzian
  manifolds. It also generalises a well respected alternative
  construction---Penrose's conformal boundary. It solves the problem of how to
  convert Penrose's definition of a gravitational singularity into a concept
  amenable to computation, by providing a location for the endpoint of the
  incomplete, inextendible, causal curve predicted by Penrose's singularity
  theorem in $\mathcal{B}(\man{M})$.

  With our semi-``completion'' of the singularity theorems, one now has the framework required to proceed to investigate whether Penrose's theorem actually predicts infinite curvature singularities like the ones found at the heart of black holes. This brings us much closer to answering the fundamental question ``What actually happens when you approach a gravitational singularity?''.

  \bibliographystyle{unsrt}
  \bibliography{Thesis-main}

\end{document}